\documentclass[runningheads]{llncs}
\usepackage{graphicx}
\usepackage{amssymb}
\begin{document}
\title{Causal Inference for Early Detection of Pathogenic Social Media Accounts
	}
%
%
\author{Hamidreza Alvari \and 
Paulo Shakarian
}
\authorrunning{H. Alvari et al.}
%
\institute{Arizona State University, Tempe, AZ, USA \\
\email{\{halvari,shak\}@asu.edu}\\
}
\maketitle              
\begin{abstract}
Pathogenic social media accounts such as terrorist supporters exploit communities of supporters for conducting attacks on social media. Early detection of PSM accounts is crucial as they are likely to be \textit{key} users in making a harmful message ``viral". This paper overviews my recent doctoral work on utilizing causal inference to identify PSM accounts within a short time frame around their activity. The proposed scheme (1) assigns time-decay causality scores to users, (2) applies a community detection-based algorithm to group of users sharing similar causality scores and finally (3) deploys a classification algorithm to classify accounts. Unlike existing techniques that require network structure, cascade path, or content, our scheme relies solely on action log of users. 

\end{abstract}
\vspace{-15pt}
\section{Introduction}
The unregulated nature of the Web and social media and the increasing privacy concerns of users mandate authorities to propose capabilities for making the Internet a safer place~\cite{Beigi2018SecuringSM,Alvari2017HT}. Countering Pathogenic Social Media (PSM) accounts (e.g. terrorist supporters~\cite{khader2016combating}) is just one instance of fighting against harmful threats to social media firms and public. They spread harmful misinformation and thus their early detection is critical as they are likely to be \textit{key} users to the formation of malicious campaigns. This is challenging because (1) social media firms manually shut down them based on reports they receive from their own users which is not always effective, (2) available data is imbalanced and network structure which is at the core of many techniques~\cite{halvarisnam2016,Beigi2018congruity} is not always available, and (3) PSMs often seek to cultivate large number of communities of supporters to spread as much harmful information as they can.

To address the challenges, causal inference is tailored to identify PSM accounts as they appear to be key users in making a harmful message ``viral''. We propose \textit{time-decay} causal metrics to distinguish PSMs from normal users within a \textit{short} time around their activity. We then investigate community structure aspect of causality for further classification improvement, by answering the following research question: \textit{Are causality scores of users within a community higher than those across different communities?} We propose a causal community detection-based classification method that takes causality scores of users and the community structure of their action log for early detection of PSMs.
\vspace{-12pt}
\section{Proposed Framework}
\vspace{-10pt}
We adopt the causal framework proposed in~\cite{DBLP:journals/corr/abs-1205-2634} and the Prima Facie notion~\cite{suppes1970} for the problem of early identification of PSMs. We extend the Kleinberg-Mishra causality ($\epsilon_{K\&M}$)~\cite{DBLP:journals/corr/abs-1205-2634} to a series of causal metrics.  Kleinberg-Mishra causality for user $i$ is computed as follows:

\begin{equation}\small
\epsilon_{K\&M}(i)=\frac{ \sum_{j\in \mathbf{R}(i)}(p_{i,j}-p_{\neg i,j}) }{|\mathbf{R}(i)|}
\end{equation}  

\noindent where $p_{i,j}$ is the probability that key users $i$ and $j$ tweet/retweet a message $m$ chronologically and make it viral. Also, $p_{\neg i,j}$ is the probability that \textit{only} key user $j$ has made a message $m$ viral by tweeting/retweeting it, i.e. user $i$ has not posted $m$ or does not precede $j$. Kleinberg-Mishra measures how causal user $i$ is by taking the average of $p_{i,j}-p_{\neg i,j}$ over $i$'s related users $\mathbf{R}(i)$. The intuition here is user $i$ is more likely to be cause of message $m$ to become viral than user $j$, if $p_{i,j}-p_{\neg i,j} > 0$. 
\vspace{-12pt}
\subsection{Leveraging Temporal Aspects of Causality}
The above metric does not capture time-decay effect and assumes a steady trend while computing causality of users. This is unrealistic, as causal effects of users may change over time. We introduce a generic decay-based metric to address the time-decay effect by assigning different weights to different time points of a given time interval, inversely proportional to their distance from $t$. This metric performs the following: (1) breaks down the given time interval into shorter time periods, using a sliding time window, (2) deploys an exponential decay function of the form $f(x)=e^{-\alpha x}$ to account for the time-decay effect, and (3) takes average of the causality values computed using the sliding time window:   

\begin{equation}\small
\xi_{K\&M}^I(i)=\frac{1}{|\mathcal{T}|}\sum_{t'\in \mathcal{T}}e^{-\sigma (t-t')}\times\epsilon_{K\&M}(i)
\end{equation}

Here, $\sigma$ is a scaling parameter of the exponential decay function, $\mathcal{T}=\{t'|t'=t_0+j\times \delta, j\in\mathbb{N} \wedge t'\leq t-\delta\}$ is a sequence of sliding-time windows, $\delta$ is a small fixed amount of time, which is used as the length of each sliding-time window $\Delta=[t'-\delta, t']$. Note that further details are eliminated due to the space limit. We have the following problem to seek if causality could be leveraged for early identifying PSMs. Note that each user is now represented by a vector $\mathbf{x}\in \mathbb{R}^d$:
\begin{description}
	\item[Problem 1 (Early Identification of PSM Accounts)] \textit{Given an action log $\mathbf{A}$, and user $u$ where $\exists t$ \textit{s.t.} $(u,m,t)\in\mathbf{A}$, our goal is to determine if $u$'s account shall be suspended given its corresponding causality vector $\mathbf{x}$ computed within a short time frame around $t$.    
	}
\end{description}

\subsection{Leveraging Community Aspects of Causality}
To answer the question posed earlier, since network structure is not provided in our dataset, we first build a graph $\mathbf{G}$ from $\mathbf{A}$ by connecting any pairs of users who have posted same message \textit{chronologically}. We leverage the \textsc{Louvain} algorithm to find the partitions over $\mathbf{G}$ and perform the following two-sample $t$-test $H_0:v_a\geq v_b,~H_1:v_a<v_b$. The null hypothesis is: \textit{users in a given community establish weak causal relations with each other compared to the other users in other communities}. We create a vector $v_a$ by computing Euclidean distances between causality vectors $(\mathbf{x}_i,\mathbf{x}_j)$ corresponding to each pair of users $(u_i,u_j)$ who are from same community. Likewise, we construct $v_b$ by computing Euclidean distance between each user $u_i$ in a community, and a random user $u_k$ chosen from the rest of the communities. The null hypothesis is rejected at significance level $\alpha=0.01$ with the $p$-value of 4.945e-17. The answer to the question is thus positive, meaning that utilizing community structure could lead to a more accurate framework. 
\vspace{-12pt}
\section{Current Progress}
\vspace{-12pt}
We conduct experiments on an ISIS dataset (in Arabic) from Twitter (details are omitted due to the lack of space). We gauge the effectiveness and efficiency (timeliness) of the approaches. We benchmark our decay-based causality metrics against the original one by incorporating them as features in (1) supervised setting, and (2) for computing proximity scores between users in community detection-based framework. The time-decay causal metrics alone reached F1-score of 0.63 and via supervised learning identified 71\% of the PSM accounts from the first 10 days of the dataset. Further, our community detection-based framework achieved the precision of 0.84 for detecting PSM accounts within 10 days around their activity; the misclassified accounts were then detected 10 days later.
\vspace{-12pt}
\section{Future Work and Advices Sought}
\vspace{-12pt}
This work is still at the early stages and consequently there are several challenges ahead. I seek insights from consortium to overcome the challenges and extend the proposed framework to better suite the problem at hand. Future plans include exploring other community detection algorithms and other causal inference tools (e.g., Granger causality). 
\vspace{-12pt}
\section{Acknowledgment}
\vspace{-12pt}
This work is an extended abstract of an ongoing research and is supported through DoD Minerva program and AFOSR (grant FA9550-15-1-0159).

%
%
%
%

\end{document}